\documentstyle[12pt,epsfig]{article} 
\def\gsim{\mathrel{\rlap {\raise.5ex\hbox{$ > $}}
{\lower.5ex\hbox{$\sim$}}}}
\def\lsim{\mathrel{\rlap {\raise.5ex\hbox{$ < $}}
{\lower.5ex\hbox{$\sim$}}}}

\newcommand{\be}{\begin{equation}}
\newcommand{\ee}{\end{equation}}
\newcommand{\bea}{\begin{eqnarray}}
\newcommand{\nn}{\nonumber}
\newcommand{\eea}{\end{eqnarray}}
\newcommand{\nd}[1]{/\hspace{-0.6em} #1}

\def\gappeq{\mathrel{\rlap {\raise.5ex\hbox{$>$}}
{\lower.5ex\hbox{$\sim$}}}}
 
\def\lappeq{\mathrel{\rlap{\raise.5ex\hbox{$<$}}
{\lower.5ex\hbox{$\sim$}}}}

\begin{document} 
\begin{titlepage} 

\begin{flushright} 
ACT-11/99 \\
CERN-TH/99-346 \\
CTP-TAMU-44/99 \\
gr-qc/9911055
\end{flushright} 

\vspace{0.5in} 
\begin{centering} 

{\Large {\bf  Gravitational-Recoil Effects on Fermion Propagation in
Space-Time Foam}} \\
\vspace{0.5in} 

{\bf John Ellis}$^a$, {\bf N.E. Mavromatos}$^b$,
{\bf D.V. Nanopoulos}$^{c,d,e}$ and {\bf G. Volkov}$^{a,f}$ 

\vspace{0.5in} 
{\bf Abstract}

\vspace{0.2in} 

\end{centering} 

{\small Motivated by the possible experimental opportunities to
test quantum gravity via its effects on high-energy neutrinos
propagating through space-time foam, we 
discuss how to incorporate spin structures
in our $D$-brane description of gravitational recoil effects {\it
in vacuo}. We also point to an interesting analogous condensed-matter
system. We use a suitable supersymmetrization of the
Born-Infeld action for excited $D$-brane 
gravitational backgrounds to argue that
energetic fermions may travel slower than the low-energy velocity
of light: $\delta c / c \sim - E / M$. It has been suggested that
Gamma-Ray Bursters may emit pulses of neutrinos at energies approaching
$10^{19}$~eV: these would be observable only if $M \gsim 10^{27}$~GeV.}

\begin{flushleft} 

\vspace{0.5in} 

$^a$ CERN, Theory Division, CH-1211, Geneva 23, Switzerland. \\
$^b$ Department of Physics, Wheatstone Laboratory, King's College
London, Strand, London WC2R 2LS, United Kingdom. \\
$^c$ Department of Physics, 
Texas A \& M University, College Station, 
TX~77843-4242, USA. \\
$^d$ Astroparticle Physics Group, Houston
Advanced Research Center (HARC), Mitchell Campus,
Woodlands, TX 77381, USA. \\
$^e$ Academy of Athens,
Chair of Theoretical Physics, 
Division of Natural Sciences, 28~Panepistimiou Avenue, 
Athens 10679, Greece. \\
$^f$ Institute for High-Energy Physics, RU-142284 Protvino, 
Moscow Region, Russia. \\
\end{flushleft}

\end{titlepage}

\section{Introduction}

It has recently been pointed out that the constancy of $c$, the
velocity of light, can be tested stringently using
distant astrophysical sources that emit pulses of radiation, such
as Gamma-Ray Bursters (GRBs)~\cite{amelino98,schaefer,efmmn}, 
Active Galactic Nuclei (AGNs)~\cite{amelino98,biller,schaefer} and
pulsars~\cite{crab}. So far, this idea 
has been explored by comparing the
arrival times of photons of different energies $E$ (frequencies $\nu$). 
It has been suggested~\cite{amelino98,ellis99,efmmn} 
that certain quantum theories of gravity might
cause variations in $c$ that increase with $E$ (or $\nu$), possibly
linearly: $(\delta c/ c) \sim (E / M)$, or quadratically:
$(\delta c/ c) \sim (E^2 / {\tilde M}^2)$, where $M$ or $\tilde M$
is a high mass scale characterizing quantum fluctuations in
space-time foam~\cite{wheeler,pullin,ashtekar99,ellis99}. 
Such a linear or quadratic dependence would enable 
any such conjectured quantum-gravity effects to be distinguished
easily (in principle) from the effects of conventional media on
photon propagation and the effects of a possible photon mass,
both of which would {decrease} with increasing energy. It is clear that,
in order to probe quantum-gravity effects by putting the strongest
possible lower limits on $M$ and $\tilde M$, there is a
premium on distant pulsed sources that emit quanta at the highest
available energy.

Unfortunately, from this point of view, the distance over which
high-energy photons can travel through the Universe is limited by
scattering on photons in the intergalactic medium. Therefore, one is led
to consider the emissions of other ultra-energetic particles, such as
neutrinos, protons and neutrons. These also scatter in the intergalactic
medium, resulting in energy cutoffs as functions of distance, e.g., the
Greisen-Zatsepin-Kuzmin cutoff for protons~\cite{GZK}. Because of their
low interaction cross sections, the best prospects for the
highest-energy quanta from the largest distances may be provided by
neutrinos.

As yet, no ultra-high-energy neutrinos have been detected, but the
sensitivity of neutrino telescopes is planned to increase
dramatically in the coming years~\cite{ANTARES}. There may well be diffuse
sources
such as ultra-heavy relics in the galactic halo, and one cannot
expect that all discrete sources will exhibit useful time structures.
However, calculations suggest that both GRBs and AGNs may 
be observable pulsed sources of high-energy neutrinos. If GRBs
do indeed emit pulses of neutrinos at energies up to $10^{19}$~eV,
as recently suggested~\cite{WB}, they might provide ideal opportunities
to probe quantum gravity (see also~\cite{alfaro}), since GRBs have
measurable cosmological
redshifts: $z \sim 1$, and exhibit short time structures: $\lsim 1$~s.
We return later to a discussion of the sensitivities to the
quantum-gravity parameters $M, \tilde M$ that such GRB neutrino
bursts might provide.

The bulk of this paper is devoted to a formal discussion of
the interaction of high-energy fermions with space-time foam.
We extend our previous $D$-brane model of quantum-gravitational
fluctuations {\it in vacuo}~\cite{ellis99} to include fermions, by
developing
the appropriate supersymmetric Born-Infeld (BI) effective action.
This enables us to demonstrate that a high-energy fermion
scattering off a $D$-brane defect in space-time induces a {\it linear}
deformation of the background metric $G_{0x} \sim E / M$, analogous
to that induced by a high-energy boson, {\it if gravitational
recoil effects are taken into account}. Section 2 contains a review
of our previous BI treatment of $D$-particle recoil, and points to an
interesting analogous condensed-matter system~\cite{volovik}.
Section 3
discusses the supersymmetrization of the BI action, and the consequences
for fermion
propagation are stressed in Section 4.

The potential phenomenological
implications for the observability of high-energy neutrino 
pulses~\cite{WB} from
GRBs and other sources are discussed in Section 5. This may be read
without ploughing through the earlier sections, if the reader is not
concerned with the formal underpinnings of the phenomenological analysis.
As we discuss there, high-energy neutrino pulses from GRBs could provide
sensitivity to $M \sim 10^{27}$~GeV or $\tilde M \sim 10^{19}$~GeV.
There is also an Appendix where certain group-theoretical
aspects of the breaking of
Lorentz invariance are developed.

\section{Space-Time Distortion due to $D$--Particle Recoil} 

We first review in more detail the theoretical
foundation underlying any such phenomenological probes of
quantum gravity. We have argued that 
virtual $D$
branes provide one possible model for space-time foam~\cite{ellis99}, and
that the recoil of a $D$ brane struck by a bosonic
closed-string particle would induce an energy-dependent
modification of the background metric `felt' by an
energetic quantum: $G_{0x} \sim u_x \sim E / M_D$. Here 
$u_x$ is the average recoil velocity of a generic $D$-brane excitation
of mass $M_D$ when struck by a boson, such as a photon,
moving in the $x$ direction with energy $E$.
Such a change in the background metric would clearly break Lorentz
invariance, but in a relatively simple one-dimensional manner
that is symmetric about the $x$ axis.
The remaining aspects of Lorentz symmetry
along directions transverse to the direction of
motion are preserved.

As preparation for our subsequent extension
of this discussion to propagating fermions, such as neutrinos,
we first review briefly the relevant $D$-brane formalism~\cite{dbranes}.
We consider the recoil induced by
a closed (super)string state (representing some
conventional matter particle) when it
strikes a $D$-particle defect in space-time. We assume that
the defect is very massive: $M_D = M_s /g_s$, 
where $M_s$ is the string mass scale: $M_s \equiv (\alpha ')^{-1/2}
\equiv 1 / \ell_s$,
with $\alpha '$ is the Regge slope, and $g_s$ is the string coupling,
which we assume to be weak: $g_s << 1$. 
It has been 
suggested in~\cite{kogan96,emndbrane,mavro+szabo}
that 
this $D$-particle recoil process 
be described 
by a logarithmic conformal field theory on the string world
sheet~\cite{lcft}.
Dynamics on the world sheet is described by a $\sigma$-model
formalism, and the pertinent $\sigma$-model perturbations are given by
world-sheet boundary deformations in the Neumann picture, of the form: 
\be
\frac{1}{2\pi \alpha '}  
\int _{\partial \Sigma} A_M (X^0) \partial _\tau X^M,~M=0,\dots 9 
\label{deform}
\ee
where the background 
gauge field $A_M (X^0)$ has the following structure: 
\be
A_M(X^0)=\left(A_0(X^0), -\frac{1}{2\pi\alpha '}Y^i(X^0)\right);
\qquad
   Y_i(X^0) = \left(\epsilon Y_i + u_i X^0 \right)\Theta _\epsilon (X^0) 
\label{comp}
\ee
where the suffix $0$ denotes temporal (Liouville) components,  
$Y_i,u_i,~i=0, \dots 9$ denotes the collective coordinates and
recoil velocity of the $D$-particle, and 
$\Theta_\epsilon (X^0)$ denotes a Heaviside step function
regulated by a cutoff parameter
$\epsilon \rightarrow 0^+$~\cite{kogan96}, which
is related to the world-sheet scale:
$  \epsilon^{-2} \sim {\rm Ln}|L/a|$, ensuring consistency of the 
logarithmic conformal algebra~\cite{lcft} satisfied by the pair of
operators 
(with couplings $Y_i$ and $u_i$) appearing in (\ref{comp})~\cite{kogan96}. 

The recoil velocity of the $D$ brane is given in terms of the 
energy/momentum transferred between the incident and the outgoing 
(low-energy) massless particle~\cite{emndbrane,mavro+szabo}:
\be
        u_i = \ell_s g_s (k_i^1 + k_i^2)
\label{emcons}
\ee
It can be shown~\cite{mavro+szabo} that this energy-momentum 
conservation relation survives the summation over world-sheet
topologies, and hence is an exact conservation law
of the full quantum system.
A technical point, which is however important for our
purposes here, is that the physical recoil velocity and 
string couplings are $\epsilon$-regularized
quantities~\cite{mavro+szabo}. 
In particular, the renormalized physical recoil velocity
and string coupling (denoted by overlines) are : 
\be
    {\overline u}_i = \epsilon u_i \qquad; \qquad {\overline g}_s =
\epsilon g_s 
\label{renorm}
\ee
This renormalization
becomes important after the identification of the change in the scale 
$\epsilon$ with a target-time
translation~\cite{emndbrane,mavro+szabo}, as we discuss later.
Upon this identification, the renormalization
(\ref{renorm})
yields the correct exactly-marginal $\sigma$-model couplings
for the recoil-velocity  deformation. The conservation of energy-momentum
(\ref{emcons}) implies that this deformation is of order $ E /
M_D$.

It was shown in~\cite{mavro+szabo} that 
the target-space dynamics
of the recoil phenomenon described by the logarithmic algebra
on the world sheet may be represented  in ten-dimensional target space
by a Born-Infeld (BI) Lagrangian for the gauge fields $A_M$. 
This picture should be contrasted with the 
Dirichlet picture, in which the target-space Lagrangian 
of a $Dp$ brane, although of BI type, 
is actually its world-volume Lagrangian. 
In our Neumann picture, the pertinent Lagrangian is of the form: 
\be
  {\cal L}_{BI} = \sqrt{\eta_{MN} + F_{MN} }
\label{BI}
\ee
where $\eta_{MN}$ is a flat Minkowski metric, and $F_{MN}=\partial _M A_N 
- \partial_N A_M$ is the Maxwell tensor for the gauge field  
$A_M$ (\ref{comp}) describing the recoil. 
We note for future reference that
the only non-vanishing components 
of the 
Maxwell tensor
$F_{\mu\nu}$,
corresponding to the recoiling background (\ref{comp}), are
$F_{0i}=u_i$, which represents a constant electric-field
background.

The propagation of ordinary photons 
has been discussed using this Lagrangian
in~\cite{ellis99}, which we follow here. In the Neumann 
picture, we simply add a similar perturbation to the $\sigma$ model,
but now in terms of a full-fledged quantum $U(1)$ field
$a_\mu$ with a field strength $f_{MN}(a)$.
The corresponding Lagrangian is then a 
straightforward  extension of (\ref{BI}):
\be 
{\cal L}_{BI} = \sqrt{\eta_{MN} + f_{MN} + F_{MN} }
\label{fFBI}
\ee
where, we repeat, $F_{MN}$ corresponds to the background (\ref{comp}),
and $f_{MN}$ is the Maxwell tensor of the 
dynamical photon field. 

As discussed in \cite{emndbrane,kanti98}, in addition to the
world-sheet boundary deformations, the perturbed theory describing the recoil 
of the $D$ particle also has a bulk graviton deformation,
due to the fact that recoil is not a conformal process
and hence requires Liouville dressing.
Writing the boundary operator (\ref{comp}) as a bulk world-sheet
(total derivative) operator, and taking into account~\cite{kogan96} 
that it has a world-sheet anomalous dimension 
$-\epsilon^2/2$, one dresses the bulk theory 
with a Liouville mode $\phi$~\cite{distler89}.
Then one identifies the Liouville field $\phi$ with the target 
time $X^0$~\cite{emndbrane}, which results in a 
metric $\sigma$-model perturbation 
of the form:  
\be
G_{ij} =\delta _{ij} , 
G_{00}=-1, G_{0i}=\epsilon(\epsilon Y_i + 
\epsilon {\overline u}_i t)\Theta _\epsilon (t),
~i=1, \dots D-1 
\label{metric}
\ee
The presence of the (regularized) $\Theta$ function
indicates that 
the induced space-time is piecewise continuous~\footnote{The 
important implications for non-thermal particle production
and decoherence for a spectator low-energy field theory
in such a space-time were discussed in~\cite{kanti98,emndbrane}.}.  
An important aspect of the approach
of~\cite{kogan96,emndbrane,mavro+szabo}
is the identification of the parameter $\epsilon$ with the target time,
for asymptotically
long times $t \gg 0$ after the collision: 
\be
\epsilon ^{-2} \sim t
\label{epstime}
\ee
The above relation should be understood as implying that 
the changes in both quantities coincide in the limit
$\epsilon \rightarrow 0^+, t \rightarrow \infty$. 

In view of (\ref{epstime}), one observes that 
the metric (\ref{metric}) 
becomes to leading order for $t \gg 0$:
\be
G_{ij} =\delta _{ij}, 
G_{00}=-1, G_{0i} \sim  {\overline u}_i, \qquad ~i=1, \dots D-1    
\label{metric2} 
\ee
and thus is constant in space-time. 
However, the metric depends on the energy content 
of the low-energy particle that scattered
off the $D$ particle, as a result of
momentum conservation during the recoil 
process~\cite{mavro+szabo}. 
We shall concentrate
on the flat asymptotic  metric (\ref{metric2}) 
in what follows.

The energy dependence of the metric is the 
main deviation from space-time Lorentz invariance 
induced by the $D$ particle recoil. 
As a result, the space-time group symmetry 
is reduced to rotations in the space-like plane
perpendicular to the direction of motion
and Lorentz boosts along the direction of motion, as
discussed in the Appendix. The residual group of transformations is a
subgroup of $SL(2,C)$. 
Upon diagonalization of the perturbed metric, one finds
a retardation in the propagation of
an energetic photon: $(\delta c / c) \sim (E / M_D)$.
The fact that propagation is subluminal, rather than 
superluminal, is linked to the underlying BI action
(\ref{fFBI}) for electromagnetism, which underlies the 
dynamics of massless photons
in the background of a recoiling brane~\cite{mavro+szabo,ellis99}.

To conclude this review section, we would like to 
draw a comparison between the above results
and some condensed-matter systems such as 
$d$-wave superconductors or superfluid ${}^3$He.
It was observed in~\cite{volovik} that 
relativistic fermionic quasiparticle 
excitations appear near the nodes of such systems, with a spin-triplet
pairing potential 
\be
    V_{{\vec p}, {\vec p}'} \propto {\vec p} \cdot {\vec p}'
\label{triplet}
\ee
and an energy gap function 
$\Delta ({\vec p}) \sim c p_x$ in the polar phase,
where $p_x$ denotes the momentum component along, say, the $x$ direction,
and $c$ denotes the effective `speed of light' in the problem. 
This is, in general, a function of the superflow velocity $w$: $c(w)$,
that is
determined self-consistently by solving the Schwinger-Dyson-type
equations that minimize the effective action.

This system was considered in the context of ${}^3$He 
in a container with stationary rigid walls and
a superflow velocity  $w$ taken, for simplicity, also along the $x$
direction.
The Doppler-shifted energy of the fermions in the pair-correlated
state with potential (\ref{triplet}) is given by 
\be
       E(p_x, \epsilon _{p} ) = \sqrt{\epsilon _{p} ^2 + c^2 p_x^2 }
+ wp_x ,
\label{he3}
\ee
where $\epsilon _p = (p^2 - p_F^2)/2m$, is the energy 
of the fermion in the absence of the pair correlation,
$p_F$ is the Fermi momentum
and $m$ is the mass of a Helium atom. The term $w p_x$
appearing in
the quasiparticle energy spectrum (\ref{he3}),
as a result of the motion of the superfluid,  
yields an effective off-diagonal 
(1+1)-dimensional metric $G_{\mu\nu}$  
with components
\be
   G^{00}=-1, \qquad G^{01} = w , \qquad G^{11}=c^2 - w^2
\label{effhe3metric}
\ee
The off-diagonal elements of the 
induced metric 
(\ref{effhe3metric})
are analogous to those of our metric 
(\ref{metric2}). In this analogy,
the role of the recoil velocity $\vec u$ in our 
quantum-gravitational case is played by
the superflow velocity field $w$. 

However, an important difference 
between our case and that of superfluid ${}^3$He is that,
in our case, the spatial elements of the metric
(\ref{metric2}) are free from the horizon problem
that characterizes the metric (\ref{effhe3metric}).
This arises when the
superflow velocity $w = c$, in which case the metric element $G^{11}$ 
in (\ref{effhe3metric})
crosses zero, leading to a signature change for superluminal 
flow $w > c$. In fact, as shown in~\cite{volovik} by 
an analysis of the gap equation, the superluminal flow branch 
is not stable, because it corresponds to a saddle point rather than a
minimum
of the effective action. This suggests that the intactness of the analogy
with our
problem, in which 
the BI action that governs the recoil 
dynamics~\cite{mavro+szabo} keeps the photon velocity
subluminal, may be maintained, as we now discuss~\footnote{Such
condensed-matter analogues of fermions moving 
in non-trivial space times may be a useful tool
for analyzing quantum-gravitational problems,
that might also be interesting to those working in the context of
the loop-gravity approach to quantum 
gravity~\cite{pullin,ashtekar99,alfaro}.}.

\section{Supersymmetric Born-Infeld Action}

It is not
immediately apparent from the BI action (\ref{fFBI}) that
a fermion such as a neutrino will also propagate
subluminally, and (if so) experience the same retardation
as a photon of the same energy. To see whether this is the
case, one should consider the recoil of a $D$ brane when
struck by an energetic fermion. It is the technical analysis of
this problem that is the next objective of this paper.
Because of the symmetries
of the scattering problem, one would expect any recoil to
be (on average) along the $x$ axis, with a velocity $\tilde u_x
\sim E/M_D$ as before. This in turn would induce a
modification $G_{0x}$ to the metric of form similar 
to that derived in the bosonic case, and hence a
corresponding modification of the velocity of
propagation: $\delta c / c \sim E / M_D$. 
To see this more mathematically, we now study a
supersymmetric extension of the above model.
This enables us, formally, to describe the
propagation of a photino, rather than a neutrino, but
we expect the conclusions to be the same.
The breaking of supersymmetry is an issue, because the
distortion of space-time induced by $D$-brane
recoil itself breaks supersymmetry~\cite{campbell-smith+mavr}.
Nevertheless, 
if one ignores gravity effects, supersymmetry still 
constrains the relevant dynamics, especially
the form of the boson-fermion interactions. 
For this reason, as we now
show, particles in a supermultiplet induce
identical recoil distortions to leading order.

A complete analysis should involve superstrings
and supermembranes, and an appropriate supersymmetric extension
of the analysis of~\cite{mavro+szabo}  
to a logarithmic superconformal algebra
on the world sheet would be necessary,
but this lies beyond the scope of this work.
As we now discuss, a relevant first step towards the introduction of 
fermions is to consider the scattering process
directly in target space-time, in
the heuristic context of a supersymmetric version of the 
$(d \lsim 10)$-dimensional $U(1)$ BI theory.  
We recall that 
a supersymmetric version of BI theory in flat $(d=10)$-dimensional 
Minkowski space-time
was considered in~\cite{schwarz97}, and is particularly simple: 
\be
{\cal L}_{SBI}  \sim \int d^{10}x \sqrt{-{\rm det}\left(\eta_{MN} 
+F_{MN} - 2 {\overline \lambda} \Gamma _M \partial _N \lambda 
+ {\overline \lambda} \Gamma ^P \partial_M \lambda 
{\overline \lambda} \Gamma _P \partial_N \lambda \right)}
\label{susy} 
\ee
This model was used in~\cite{schwarz97} 
to study $D$ branes in the Dirichlet picture.
In this sense, the ten-dimensional 
Lagrangian (\ref{susy}) was applied to the world volume of a nine-brane. 
In that case there were two supersymmetries,
one of which was spontaneously broken by the presence of the $D$ brane,
with the photino $\lambda$ the corresponding goldstino
particle of spontaneously-broken Poincare
symmetry.
The second supersymmetry is more subtle, but its 
appearance is explained in~\cite{schwarz97}. 

In a conventional 
string-theoretic approach, in order
to obtain the form of the $(d < 10)$-dimensional BI action
relevant for our purposes here, one needs to implement 
dimensional reduction of the above action, which leads to 
extended supersymmetries. 
However, in our approach, one may obtain directly a 
four-dimensional BI action, by choosing the 
recoil background deformation (\ref{comp}) appropriately, 
i.e., restricting oneself to 
$u_i$ with non-trivial components
{\it only} for $i=1,2,3$. 
In such a case, one may simply discuss a $N=1$ target-space 
supersymmetrization of the four-dimensional 
BI action~\cite{nunez}. 
This is what we do below, using it as a toy model for the 
discussion of fermion propagation in our 
recoiling $D$--brane framework. 

We start
with the bosonic part of the four-dimensional BI Lagrangian:
\be
L_{BI}=\beta^2\left(1 - \sqrt{-{\rm det}\left(g_{\mu\nu} 
+ \frac{1}{\beta}F_{\mu\nu} \right)}\right)
\label{4dBI}
\ee
where the signature of the metric is assumed to be $(+,---)$.
We have in four space-time dimensions the identity 
\be
{\rm det}\left(g_{\mu\nu} + \frac{1}{\beta}F_{\mu\nu}\right)
= -1 -\frac{1}{2\beta^2}F_{\mu\nu}^2 + \frac{1}{16\beta^4}\left(F_{\mu\nu}
{\tilde F}^{\mu\nu}\right)^2~, ~~{\tilde F}_{\mu\nu}=
\epsilon_{\mu\nu\rho\sigma}F^{\rho\sigma} 
\label{BIide}
\ee
which allows the BI action to be expressed in terms of quadratic structures
of the Maxwell tensor. An important ingredient is 
the appearance of the CP-violating term $F{\tilde F}$, which 
is a characteristic
feature of the gauge action in four dimensions. 
In the approach to $D$--brane recoil 
of~\cite{mavro+szabo}, the $U(1)$ field is 
a background  (\ref{comp}) associated with the 
collective coordinates of the $D$--brane soliton. 
The quantity $\beta$ is related to the string coupling and the 
string length by:
\be
    \beta = 1/(\ell_s {\overline g}_s)
\label{beta}
\ee
where ${\overline g}_s$ is the physical string coupling, renormalized 
(\ref{renorm}) in the sense of~\cite{mavro+szabo}. 

We next consider treating~\cite{ellis99}
the interactions of photons with the background of 
recoiling $D$ branes through (\ref{fFBI}),  
as appropriate for the Neumann picture~\cite{mavro+szabo}.
Due to the identity (\ref{BIide}), it is evident that {\it if one
ignores the gravity effects (\ref{metric2})}, the leading corrections
to photon propagation 
will come from Lorentz-invariant terms of the form 
$f_{\mu\nu}^2 \times {\cal O}(u_i^2)$, i.e., quadratic in 
the small recoil velocity $u_i$, and hence corrections
are suppressed by quadratic powers of $M_s$, as expected due to
Lorentz invariance. {\it Our key step is to go beyond this, by treating
gravitational recoil effects.}

For the purposes of supersymmetrization, we treat the $U(1)$ 
gauge field $A_\mu$ as a fulll-fledged quantum field, and not simply as 
a background related to the collective coordinates of the $D$ brane.
The $N=1$ supersymmetric version of (\ref{4dBI}) can be constructed 
in a compact form if one uses superfields~\cite{nunez}:
\be
    L_{BI}^{susy}=\frac{1}{4}\{ \int d^2\theta W^2 + 
\int d^2{\overline \theta} {\overline W}^2 \} + 
\sum_{s,t=0}^{\infty} a_{1st}\int d^4 \theta W^2 {\overline W}^2 
X^s Y^t 
\label{superfield}
\ee
where $W_\alpha$ is the field-strength chiral supermultiplet, 
related to the vector 
superfield $V_\alpha$ in the usual way, and 
$X,Y$ are appropriate superfields, whose bosonic components
read:
\bea
&~& 
X_{|_{\theta={\overline \theta}=0}}
= -\beta^{-2}D^2 -\frac{1}{2}\beta^{-2}F_{\mu\nu}^2 
-i\lambda \nd{\partial} {\overline \lambda} 
-i{\overline \lambda} \nd{\partial} \lambda ~, \nn \\
&~& Y_{|_{\theta={\overline \theta}=0}}= 
\frac{1}{2}\beta^{-2}F_{\mu\nu}{\tilde F}^{\mu\nu} +
\lambda \nd{\partial} {\overline \lambda} - 
{\overline \lambda} \nd{\partial} \lambda
\label{xyfields}
\eea
where $D$ is an auxilliary field and $\lambda$ the photino
field, which is a two-component Majorana spinor~\footnote{It is
interesting to note that the action (\ref{superfield}) yields
pairing interactions similar to (\ref{triplet}).}. The
expansion coefficients
$a_{1st}$ are expressed in terms of inverse powers of 
the coupling $\beta ^2$~\cite{nunez}. 
For our purposes, we note that the supersymmetric 
extension (\ref{superfield}) yields three kinds of terms: (i) 
pure bosonic terms, which yield the bosonic BI Lagrangian (\ref{4dBI})
when one uses the equations of motion for the auxiliary field $D$ (which
also yield $D=0$), 
(ii) self-interacting fermion terms $L_{f}$, and (iii) boson-fermion 
interactions, $L_{fb}$, which include the kinetic term
for the fermions. 
The latter are the terms needed
for our purposes, and we concentrate on them
henceforth.
Their detailed structure is given in \cite{nunez},
and will not be given here. It is sufficient for our purposes
of describing recoil induced by fermion scattering
to restrict ourselves to a background of the form (\ref{comp}),
whilst keeping the photino field a full-fledged quantum field.

Combining the two-component fermions 
$\lambda_\alpha, {\overline \lambda}_{\dot \alpha}$
into a four-component Majorana spinor 
\be
\Lambda = \left(\begin{array}{c} \lambda _\alpha \\ {\overline \lambda}_{\dot
\alpha} \end{array}\right)
\label{majoron}
\ee
one observes that 
the relevant $N=1$ supersymmetry transformations can be 
expressed in the form:
\bea 
&~&\delta^S A_\mu =-i{\overline \varepsilon}\gamma _\mu \Lambda, \qquad
\delta^S \Lambda -i\left(\Sigma^{\mu\nu}F_{\mu\nu} + \gamma^5
D\right)\varepsilon , \nn \\
&~& \delta^S D=i{\overline \varepsilon}\gamma^5 \nd{\partial}\Lambda 
\label{susytrans}
\eea
where the upper index $S$ in $\delta^S$ denotes a supersymmetry transformation,
$\varepsilon$ is the appropriate (infinitesimal) supersymmetry parameter, 
$\Sigma_{\mu\nu} \equiv \frac{i}{4}[\gamma_\mu, 
\gamma_\nu]$, and $\gamma^5 \equiv i\gamma^1\gamma^2\gamma^3\gamma^0$.

We are now in a position to discuss the compatibility of the 
background (\ref{comp}) with $N=1$ supersymmetry. 
As is obvious from the form of the supersymmetry transformation
(\ref{susytrans}) and from the form of the bosonic background
(\ref{comp}),
compatibility with supersymmetry 
can be achieved for `photino' fields $\Lambda$ 
which are independent of space and depend linearly on time $X^0$.
A generic form for the Majorana spinor $\Lambda$ 
would then be:
\be
     \Lambda = \epsilon \Lambda_1 + \Lambda_2 x^0
\label{spinors}
\ee
The quantities $\Lambda_{i}$,$i=1,2$ are quantized as a result 
of the summation over genera in a 
world-sheet framework~\cite{emndbrane,mavro+szabo}.
Although, rigorously, one should first explicitly check that 
supersymmetry survives such a resummation over higher
world-sheet topologies, i.e., there are no anomalies associated with its
quantization, here
we simply assume this is the case. The $N=1$ supersymmetry transformation
(\ref{susytrans}) would then imply: 
\bea
&~&     \delta ^S Y_i =-i{\overline \varepsilon} \gamma_i \Lambda_1, 
\qquad \delta ^S u_i =-i{\overline \varepsilon} \gamma_i \Lambda_2~, \nn
\\
&~& \epsilon \delta^S \Lambda_1 + \delta^S \lambda_2 X^0 =-i\Sigma^{0i}u_i \varepsilon
+ i\gamma^5 D \varepsilon, 
\qquad \delta ^S D=i{\overline \varepsilon} \gamma^5 \gamma^0\partial_0
\Lambda_2~,
\label{comptrans}
\eea
from which it is clear that the $N=1$ supersymmetric partner 
of the background (\ref{comp}) is the one with $\Lambda_2=0$, implying
$D=0$ and $\delta^2 D= \delta^S u_i = 0$, 
which is compatible with on-shell supersymmetry~\footnote{We recall that
the background 
(\ref{comp}) is a solution of the classical equations of motion.}. 
Thus, in flat target space times, the background is
compatible with $N=1$ supersymmetry. 

We now study spinor propagation 
in the background (\ref{comp}). We notice first that this background
conserves CP, since  ${\tilde F}_{\mu\nu}=0$, and then 
make a derivative expansion of the fermion-boson interactions.
Restricting ourselves to the leading order in this expansion,
we obtain the terms: 
\be
   L_{fb} \ni -\frac{i}{2}{\overline \Lambda}\nd{\partial}\Lambda 
-\frac{i}{8}{\overline \Lambda}\nd{\partial}\Lambda \left(D^2 
+ \frac{1}{2}F_{\mu\nu}^2\right) -\frac{i}{4} {\overline \Lambda}
\gamma^\mu \partial^\nu \Lambda F_{\nu\rho}F^{\rho}_\mu + \dots 
\label{lfb}
\ee
where the $\gamma^\mu$ are $4 \times 4 $ Dirac  matrices, and 
the $\dots$ denote subleading derivative terms.

Using the background (\ref{comp}), then, 
we obtain from (\ref{lfb}):
\be
   L_{fb} \ni -\frac{i}{2}\left(1 - 
\frac{1}{8}u_i^2\right)
{\overline \Lambda}\nd{\partial}\Lambda 
+\frac{i}{4} u_iu_j {\overline \Lambda}
\gamma^i \partial^j \Lambda 
- \frac{i}{4} u_i^2 {\overline \Lambda}
\gamma^0 \partial^0 \Lambda 
 + \dots 
\label{lfb2}
\ee
We observe that, in flat space-times,
supersymmetrization of the BI action implies
non-trivial propagation of the
massless `photino' field in the recoil background 
(\ref{comp}). Moreover, as with the bosonic counterparts, 
the effects are suppressed
by quadratic inverse powers of $M_D=M_s/g_s$,
with $M_s = \ell_s^{-1}$. This may be traced
back to the Lorentz-invariant form of the flat-space
BI action (\ref{4dBI}),(\ref{BIide}). 
As we discuss below, it is only after 
coupling to gravity, which manifestly breaks Lorentz invariance, 
that the modification of the propagation becomes linear.
Such linear terms arise from the kinetic term of the photino
field after coupling to gravitational backgrounds. 

Before analyzing this issue, we first comment
on the extended supersymmetries that characterize
BI actions when the latter are
viewed as world-volume actions of $D3$
branes, which is different from 
the picture described above. 
Such supersymemtric formulations
are obtained by appropriate dimensional reductions of the 
ten-dimensional flat Minkowski space-time.
Six of the coordinates give rise to scalar fields,
$y^i, i=4,5,\dots 9$ in  the four-dimensional world-volume
theory.
In that case there is an extended supersymmetry of the $N=4$
Yang-Mills type, as discussed in \cite{shmakova}.
The spectrum of the 
gauge-fixed supersymmetric formulation of the $D3$-brane action
in a flat space-time background consists of
the world-volume Abelian gauge field
$A_\mu , \mu=0 \dots 3 $, four four-component $d=4$ Majorana spinors
(extended `photinos') $\Lambda_\alpha ^I, I=1 \dots 4$, where $\alpha$
is a superfield spinor index:
\be
\Lambda^I = 2\left(\begin{array}{c} \lambda ^I_\alpha \\ 
{\overline \lambda}_{I,\dot
\alpha} \end{array}\right)
\label{majoron2}
\ee     
and the scalar fields obtained from the dimensional reduction
of the ten-dimensional theory, which are
conveniently written as $s^{IJ}=\frac{1}{2} 
\left({\tilde \sigma}_{t}\right)^{IJ}y^t, I,J=1, \dots 4 , t=4, \dots, 9$,
where the ${\tilde \sigma _t}$ are $ 4 \times 4$ 
matrices appearing in the chiral 
representation of the Dirac matrices in six dimensions. 
In this way, there is a manifest $SU(4)$ symmetry, which makes the problem
analogous to $N=4$ supersymmetric Yang-Mills theory, in terms of 
expressed in terms of $d=4$-dimensional `Yang-Mills' variables 
$(A_\mu, \Lambda^I, s^{IJ} )$. 

The above construction is potentially useful, in that it
incorporates four species of 
Majorana fermions, including those that become members of
chiral supermultiplets when $N=4$ supersymmetry is eventually broken
down to $N=1$. Hence it may be
closer to providing a
toy model for neutrino propagation in space-time foam.

The formalism of \cite{mavro+szabo} applies intact 
to the description of the recoil of the $D$ particle after
scattering by low-energy supersymmetric matter, i.e., photons and
photinos, on the world-volume $D3$ brane
The recoil appears as a background contribution
to the four-dimensional world-volume gauge potential of the form (\ref{comp}).
The pertinent interaction terms can be read easily from the 
component form of the supersymemtric Lagrangian given in 
\cite{shmakova}, and again, one arrives at similar conclusions
(\ref{lfb2}) as above.
The advantage of the above world-volume 
formalism is that one may combine
two Majorana neutrino species into a Dirac one, and thus discuss
formally the propagation of massless Dirac spinors as well.

\section{Fermion Propagation in a Space-Time Metric Distorted by
Gravitational Recoil}

So far, using the recoil formalism of~\cite{mavro+szabo}, 
and the appropriate supersymmetrization of the 
BI lagrangian, we have discussed the propagation of $U(1)$
vector particles and the corresponding `photinos',
{\it ignoring the effect of gravitational recoil on the background}.
In this simplified case, the velocities of both photons
and photinos differ from the naive low-energy value $c$
by amounts that are suppressed quadratically
by two inverse powers of the string or $D$--brane scale, which are
assumed in four-dimensional models to be near the Planck scale
$\sim 10^{19}$ GeV. Now it is time to explore the
effect on fermions of the distorted gravitational background given
by the metric (\ref{metric2}). 

As was discussed in section 2,
the appearance of such a metric has been proven in the 
bosonic part of the world-sheet $\sigma$ model 
of the string, using an appropriate 
Liouville dressing on the world-sheet, and identifying the 
Liouville field as the target time, as explained in~\cite{emndbrane}. 
A similar procedure should be valid for superstrings, providing a formal
arguments that fermions and bosons should create similar 
metric backgrounds when they scatter off a $D$ particle.
A complete proof of this would involve extending
the Liouville analysis of~\cite{emndbrane}
to a world-sheet superfield Liouville formalism.
We do not present such a proof here, but limit ourselves to
heuristic arguments why fermions should induce a modified
metric analogous to (\ref{metric2}).

Physically, one expects a high-energy incident fermion to
induce a $D$-brane recoil which is similar at least parametrically
to that induced by an incident boson, since the most important
kinematic constraint is that of energy-momentum conservation
(\ref{emcons}). Just as in the bosonic case, the $D$-brane
recoil velocity $\tilde u_i$ should be of order $E / M$, where
$M$ is of order the Planck or string scale. The only possible
difference might be in the angular distribution of the recoil
induced by fermion scattering. This order-of-magnitude argument
would be strengthened in the limit of supersymmetry. As
mentioned earlier, the recoil process itself violates
supersymmetry, e.g., because it causes a deviation from the
ground-state energy. However, we expect this breaking of
supersymmetry to be negligible at high energies. In any case,
we know that supersymmetry is not exact even in the ground state,
so any argument based on exact supersymmetry should be
treated with caution, except at high energies much larger
than the supersymmetric mass splitting. This is actually the case
for the main application we make at the end of this paper, namely
to fermions with energies approaching $10^{19}$~eV. However, even
at lower energies we expect the basic kinematic argument
concerning the magnitude of the recoil velocity to be valid.
Since the metric perturbation (\ref{metric2}) is directly
related to this recoil velocity, we also assume that the
metric deformation induced by an energetic fermion is also of the generic
form (\ref{metric2}).

The next step is to consider the velocity of fermion
propagation in such a deformed metric.
In \cite{efmmn} there is a simple description of the propagation of 
electromagnetic waves in such a background, and 
the corresponding induced refractive index, based on
an elementary analysis of Maxwell's equations. We now
carry out a similar analysis using the massless Dirac
equation to calculate the fermion propagation.

The Dirac equation in an external gravitational field can be written 
using the spin connection given by Fock-Ivanenko coefficients:
\begin{eqnarray}
\Gamma_{\mu}\,=
\, \frac{i}{4}\cdot \gamma^{\nu} \cdot \gamma_{\nu;\,\mu}\,
=\,\,-\frac{1}{4} \cdot e^{\nu}_m \cdot e_{\nu\,n;\,\mu} \cdot 
\sigma^{m\,n}
\end{eqnarray}
where $ \sigma^{m\,n}\, \equiv \,-\frac{1}{2} \cdot
 [\tilde\gamma^m,\tilde\gamma^n]_{-},$
where we use the usual relations between the general relativity
$\gamma^{\nu}$ and 
Lorentz ${\tilde\gamma}^m$ matrices:
\begin{eqnarray}
\gamma^{\nu}\,&=&\, e^{\nu}_m \cdot {\tilde\gamma}^m, \nonumber\\
\gamma_{\nu}\,&=&\, e^m_{\nu} \cdot {\tilde\gamma}_m 
\end{eqnarray}
and 
\begin{equation}
\{\gamma^{\mu},\gamma^{\nu}\}_{+}\,=\,
 2 \cdot    g^{\mu\,\nu}, \,\,\,
\{\tilde\gamma^m, \tilde\gamma^n\}_{+}\,=\,
 2 \cdot \eta^{m\,n},
\end{equation}
as usual.

Assuming the small metric perturbation 
(\ref{metric2}), about flat Minkowski space time, with $|\vec u| <<1$,  
one has get  the following expression for the
vierbeins:
\begin{equation}
e^{\nu}_m\, = \, e^{m}_{\nu}\, = \pmatrix {
-1&0&0&-u_1 \cr
0&-1&0&-u_2 \cr
0&0&-1&-u_3 \cr
0&0&0&1  \cr }
\label{vierbein}
\end{equation}
The general form of the Dirac equation can be written in the next form:
\begin{equation}
\{  \gamma^{\nu} \cdot (\nabla_{\nu} \,-\,\Gamma_{\nu}) \} \psi\,=\,0
\label{newDirac}
\end{equation}
where the covariant derivative $\nabla_{\nu}$ is derived from the
connection
$\Gamma^{\alpha}_{\mu \,\nu}$:
\begin{equation}
\nabla_{\nu}\,=\, \partial_{\nu}\,+\,\Gamma_{\nu \,\mu}^{\mu}
\end{equation}
In our metric (\ref{metric2}), the  Dirac equation (\ref{newDirac})
becomes
\begin{equation}
{ \{ ({\tilde\gamma}^{m} \cdot \partial_{m} )\, 
-\,\tilde\gamma^0 \cdot (\vec{u} \cdot \vec{\nabla})\,\} \Psi \,=\,0.}
\label{modDirac})
\end{equation}
In order to derive the dispersion relation for the massless fermion,
we act on (\ref{modDirac}) from the left with
the operator $\tilde\gamma^{\mu} \cdot \partial_{\mu}$
to obtain:
\begin{equation}
\{{\partial_0}^2\,-\,{\vec\nabla}^2\,-
\,2\cdot (\vec u \cdot \vec \nabla) \partial_0 \} \cdot \Psi\,=\,0. 
\end{equation} 
It is then easy to see that the correct dispersion relation between the
energy $E$ and momentum $p$ of
the massless fermion in the metric background
(\ref{metric2}) is:
\begin{equation}
{E}^2\,=\,{p}^2\,-\,2 \cdot  E\cdot (\vec p \cdot \vec u): \qquad
| \vec u |~ \sim E / M.
\label{dispferm}
\end{equation}
which is similar to that obtained previously for a massless
boson (photon), as we expected.

The naturalness of this conclusion can also be seen by
considering the group theory of the perturbation (\ref{metric2}), as
discussed in the Appendix. 
The metric is invariant under a
subgroup 
of the ${SL(2,C)}$ transformations
which leave the magnitude of the vector ${\vec u}$ invariant,
so one expects the dispersion relation to have the same
invariance properties, which leaves (\ref{dispferm})
as the unique possibility.

We also comment that similar
quantum-gravity corrections 
to the dispersion relations of bosons and fermions
in space-time foamy backgrounds 
has also been observed in the loop approach to quantum-gravitational 
space-time foam, an a study of massive
spin 1/2 fields~\cite{alfaro}. The corrections in that case 
result~\cite{pullin} from the discrete (cellular) structure of space-time
at Planckian distances, which is a characteristic feature
of the loop-gravity approach~\cite{ashtekar99}. 
The temptation to take (\ref{dispferm}) as a serious
possibility and explore its phenomenological consequences
can only be enhanced by this convergence of two very
different approaches to the description of space-time foam.

\section{The Phenomenology of High-Energy Neutrino Pulses}

We now apologize to any astrophysical readers for the
previous formal excursion,
and now consider the possible observational implications
of the modification (\ref{dispferm}) of the propagation of massless
fermions in the gravitational background (\ref{metric2}) induced
by gravitational recoil: $| \vec u | \sim E / M$. 

As we have pointed out previously~\cite{amelino98,efmmn} in analyses of
possible deviations
of photon velocities from the naive velocity of light, 
the differences in travel times $t$ of particles with
different energies are given to leading order by
\begin{equation}
\delta t \sim - L \cdot \delta c
\label{delay}
\end{equation}
in units where $c = 1$, where $\delta c \sim - E / M$, resulting in
\begin{equation}
\delta t \sim L \cdot E / M
\label{FofM}
\end{equation}
so there is a premium on observing astrophysical sources 
at large distances $L$ that emit high-energy pulses with narrow time
structures, so as to be sensitive to the largest value of
$M$~\footnote{As in the case of photons, our medium effects can
easily be distinguished from those of a neutrino mass~\cite{Stod}.}.

Possible examples might include AGNs, at typical redshifts
$z \sim 0.03$, and GRBs at $z \sim 1$, though we emphasize that
no energetic neutrinos from such distant sources have yet been
observed.
Concerning AGNs, neutrino energies comparable to the maximum 
observed $\gamma$ energies of around $10^{12}$~eV could be
envisaged, whereas it has recently been proposed that GRBs might
emit neutrinos with energies as high as $10^{19}$~eV~\cite{WB}.
AGN $\gamma$-ray fluxes are known to exhibit time variations on
scales down to $300$~s, and
one might conjecture a similar time scale for possible
$\nu$ emissions. On the other hand, a typical GRB time scale is
1~s, and even much shorter time scales have been observed, though
not (yet) in the subsample of GRBs known to have cosmological redshifts.

Because of their similar $\nu$ and $\gamma$ energies,
the sensitivity to deviations from the low-energy
velocity of light $c$ obtained from observations of AGN
neutrinos would be comparable to that from photons, 
namely approaching $M \sim 10^{17}$~GeV, as discussed
elsewhere~\cite{biller}. On the
other hand, to estimate the possible sensitivity that could be obtained from 
an observation of a GRB $\nu$ pulse, we assume a
distance of 3000~Mpc, $\nu$ energies $E \sim 10^{19}$~eV and a pulse
resolution of 3~s, leading to a sensitivity to $M \sim 10^{27}$~GeV!
This is many orders of magnitude beyond what could be achieved with
photons, because of their much shorter mean free path at high energies,
and is far beyond the normal Planck scale $M_P \sim 10^{19}$~GeV. 
If our proposal of a linear deviation $\delta c \sim - E / M$ of the
velocities of
high-energy particles from the canonical low-energy velocity of
light $c$, with $M \lsim M_P$ is correct, {\it no such pulse
should ever be seen}. Conversely, if such
a high-energy neutrino pulse were to be seen, it would cast doubt on
our expectation of a linear deviation of the velocities of
high-energy particles from the canonical low-energy velocity of
light $c$~\footnote{For the record, we comment that even if the
deviation from $c$ were only quadratic: $\delta c \sim - (E / \tilde
M)^2$, which we repeat that we do not expect, such a high-energy GRB $\nu$
pulse would be sensitive to $ \tilde M \sim 10^{19}$~GeV!}. 

\section{Summary and Prospects}

High-energy neutrinos from astrophysical sources may offer
an unparallelled observational opportunity to study whether
the velocity of light is universal, since they can propagate
across the entire universe essentially unimpeded.
Encouraged by this possibility, we have given in this paper
heuristic arguments extending our previous suggestion
that $\delta c \sim - E /M$ for energetic photons to the
analogous case of fermions. Our arguments have
been based on a supersymmetric extension of our previous $D$-brane
approach to modelling the medium properties of the
quantum-gravitational vacuum, and on an analysis of
fermion propagation in the perturbed gravitational
background metric (\ref{metric2}) suggested by these
arguments. We re-emphasize that the suggested linear effect
would be due to {\it gravitational recoil}, and that the
neglect of this possibility would lead to a quadratic
deviation $\delta c \sim - (E / \tilde M)^2$. A linear
deviation has also been motivated by studies within the
loop approach to quantum gravity~\cite{alfaro}.

Much work remains to be done, even within the framework
of the $D$-brane approach espoused here. For example, 
the development of the appropriate logarithmic superconformal algebra
would be of formal interest. More practically, it
would be good to rederive the gravitational-recoil effect
for fermions without appealing at all to supersymmetry, which is
certainly broken, both in the ground state and by the recoil
process itself~\cite{campbell-smith+mavr}. At a more profound level, it
is desirable to
establish more firmly the theoretical foundations of the $D$-brane
approach, and to relate it more directly to alternatives such as
the loop approach to quantum gravity.

Despite these theoretical lacunae, we believe that we have
provided in Sections 2 to 4 sufficient motivation from
fundamental physics to take an active interest in the
observational opportunity that may be provided by distant
high-energy neutrino sources. Moreover, as seen in Section 5,
the sensitivity these could offer to possible deviations of
high-energy particle velocities from the canonical low-energy
velocity of light are very impressive: plausible GRB parameters
could provide sensitivity to $M \sim 10^{27}$~GeV in
the (favoured) case of a linear dependence on energy, and even
$\tilde M \sim 10^{19}$~GeV in the (non-gravitational) case of
a quadratic dependence. Let us hope that Nature obliges us by 
providing these or other
such distant and pulsed sources of high-energy neutrinos.

\section*{Appendix: Reduced Lorentz Symmetry}

We recall that there is a homomorphism: ${A \rightarrow L_A}$
which relates elements $A$ of the universal covering group
${SL(2,C)}$ to elements $L_A$
of the connected component ${L_{+}^{\uparrow}}$ of the Lorentz
group $L$ with positive determinant and time-like coefficients
$L^0_0$. We also recall that any ${A\in SL(2C)}$  can be
decomposed as
\begin{eqnarray}
A\,=\,B \cdot \Omega, 
\end{eqnarray}
where the Hermitian matrix ${B}$ and the unitary matrix ${\Omega}$
determine the  boost and the 3-space rotation, respectively:
\begin{eqnarray}
B\,&\,=\,&\, \exp{(\omega/2)  (\vec{n} \cdot \vec {\sigma})},\nonumber\\
\Omega\,&\,=\,&\, \exp{(-i \cdot \varphi/2)  (\vec{n} \cdot \vec {\sigma})}.
\end{eqnarray}
where $\vec{\sigma}$ denotes the $2 \times 2$ Pauli spin matrices, and
the operator ${L_{B}}$ is a pure ${\vec{n}}$-boost with a velocity
\begin{equation}
\frac{\vec{V}}{c}\,=\,\tanh{(\omega)} \cdot \vec{n}.
\end{equation}
For example, for a pure Lorentz boost in the direction ${x_1}$ 
the corresponding  matrix ${A \in SL(2,C)}$
has the following simple form:
${A\,=\, a_0 \cdot \sigma_0\,+\, a_1 \cdot \sigma_1}$ 
with the following constraint on the real parameters: ${a_0=\cosh
{\omega/2}}$ and 
${a_1=\sinh {\omega/2}}$, such that ${a_0^2\,-\,a_1^2\,=\,1}$. 

Let us now consider the interval in our modified metric (\ref{metric2}):
\begin{eqnarray}
d s^2\,=\,g^{\mu\,\nu}\cdot dx_{\mu}  dx_{\nu}\,=\,
-\,d x_1^2\,-\,dx_2^2\,-\,dx_3^2\,+\,dx_0^2\,-
\,2 \cdot \vec{u} \cdot \vec{dx} \cdot dx_0,
\end{eqnarray}
where
\begin{equation}
g^{\mu\nu}\,=\,\pmatrix{
-1&0&0&-u_1\cr
0&-1&0&-u_2\cr
0&0&-1&-u_3\cr
-u_1&-u_2&-u_3&1\cr}
\end{equation}
and $x_0=c \cdot t\,=-i\cdot x_4$.

One can introduce a group of transformations 
that leave this metric invariant:
\begin{equation}
{ \hat\Lambda\,=\,\hat O^{-1} \cdot L  \cdot {\hat O}},
\end{equation}
where $ { L}$ is an ordinary Lorentz transformation, such as a boost
$\omega$ in the $x$ direction. 
The operator ${\hat O}$ diagonalizes
the metric in the form:
\begin{equation}
{g^{D}_{m \,n}\,=\,diag(-\,r^2,\,-\,1,\,-\,1,\,+\,r^2),}
\end{equation}
where $ r\,=\,\sqrt{(1+{\vec u}^2)}$.
In the special case,  $ u_1 \not =1,
u_2=u_3=0$ and  $ r=\sqrt{1+u_1^2}$, one finds:
\begin{equation}
O\,= \, \pmatrix{
\frac{1}{r} &0&0& \frac{\sqrt{r^2-1}}{r}\cr
0&1&0&0\cr
0&0&1&0\cr
0&0&0&1\cr}
\end{equation}
and the general case is also easily found.

It is also possible to define transformed Pauli matrices:
\begin{equation}
\hat {\sigma}(x)\,=\,\pmatrix{
x_0 -\sqrt{r^2-1} \cdot x_1+x_3& r \cdot x_1-ix_2\cr
 r \cdot x_1+ix_2&x_0 -\sqrt{r^2-1} \cdot x_1-x_3\cr},
\end{equation}
whose determinant reproduces the deformed metric, where
\begin{eqnarray}
\hat{\sigma}_i\,=\,\sigma_i,\,\,\,\, i=0,2,3
\end{eqnarray}
and
\begin{equation}
\hat {\sigma_1}\,=\,r \cdot \sigma_1\,-
\,\sqrt{r^2-1} \cdot \sigma_0\,=\,\pmatrix{
-\sinh {\varpi}&\cosh {\varpi}\cr
\cosh {\varpi}&-\sinh {\varpi}\cr}
\end{equation}
where $\cosh {\varpi}=r$ and  $\sinh {\varpi}=\sqrt{r^2-1}$.

The modified Lorentz transformations ${\hat{\Lambda}}$ 
can be  expressed by the following
$SL(2, C)$ matrices:
\begin{eqnarray}
\hat{\Lambda}_{0\,n}\,&=&\,\frac{1}{2} \cdot 
\{Tr ( A \hat{\sigma}_n A^{+} )\,+\, \tanh {\varpi} 
Tr ({\sigma}_1 A \hat{\sigma}_n A^{+} )\}\nonumber\\
\hat{\Lambda}_{1\,n}\,&=&\,\frac{1}{2} \cdot \frac{1}{\cosh{ \varpi}}
 \cdot \{Tr ({\sigma}_1 A \hat{\sigma}_n A^{+} )\}\nonumber\\
\hat{\Lambda}_{2\,n}\,&=&\,\frac{1}{2} \cdot 
\{Tr ({\sigma}_2 A \hat{\sigma}_n A^{+} )\},\nonumber\\ 
\hat{\Lambda}_{3\,n}\,&=&\,\frac{1}{2} \cdot 
\{Tr ({\sigma}_3 A \hat{\sigma}_n A^{+} )\}. 
\end{eqnarray}
A boost in the ${x_1}$ direction  
can be expressed by the following type of modified
symmetric and unimodular Lorentz operator:
 \begin{equation}
{\hat\Lambda_{B_1}}^T\,=\,\hat\Lambda_{B_1}\,=\, 
{O}^{-1} \cdot  {L}_{B_1} \cdot  O,\,\,
\,\,\,\,\,\,\,\, det (\hat\Lambda_{B_1})=1. 
\end{equation} 
which forms a one-parameter subgroup.
The composition of two boosts ${\Lambda_{B_1}(\omega_1| \varpi)}$ and
${\Lambda_{B_1}(\omega_2| \varpi)}$, 
is itself a boost ${\Lambda_{B_1}(\omega| \varpi)}$
in the same direction, with parameter  ${\omega =\omega_1 +
\omega_2}$:
\begin{equation}
\Lambda_{B_1}(\omega_1| \varpi)\cdot 
\Lambda_{B_1}(\omega_2| \varpi)\,=\,
\Lambda_{B_1}(\omega_1+\omega_2| \varpi).
\end{equation} 
There is also a one-parameter group symmetry
connected to pure rotations in the ${(x_2-x_3)}$ plane:
\begin{equation}
\hat\Lambda_{\Omega_{23}}\,=\, 
O^{-1} \cdot L_{\Omega_{23}} \cdot O.
\end{equation}
Thus, for finite $|u|$, there exists a two-parameter 
symmetry group, generated by the above boosts and rotations:
\begin{equation}
\hat {\Lambda}\,=\,
\hat{\Lambda}_{B_{\vec {u}}}\cdot \hat{\Lambda}_{{\Omega}_{\vec {u}}}
\end{equation} 
surviving from the full $SL(2,C)$ group.

\section*{Acknowledgements} 

The work of N.E.M. is partially supported by P.P.A.R.C. (U.K.), that of
D.V.N. by D.O.E. grant DE-F-G03-95-ER-40917, and that of G.V. by the
World Laboratory. N.E.M. and D.V.N. thank Hans Hofer 
and G.V. thanks A. Zichichi for interest and support. N.E.M. and G.V.
also thank the CERN Theory Division for its support.

\end{document}